# THEORETICAL STUDY OF QUANTUM DISSIPATION AND LASER NOISE EFFECTS ON THE ATOMIC RESPONSE

O. El Akramine♣, A. Makhoute♣*, M. Zitane♣ and M. Tij♣

♣ UFR Physique Atomique, Moléculaire & Optique Appliquée
Université Moulay Ismail, Faculté des Sciences,
B.P. 4010 Beni M'hamed, Meknès, Morocco.

\* Physique Atomique Théorique, Faculté des Sciences,
Université Libre de Bruxelles,
CP 227, Brussels, Belgium.

# Abstract

The nonlinear dynamics of dissipative quantum systems in incoherent laser fields is studied in the framework of master equation with random telegraph model describing the laser noise and Markovian approximation for dealing with the system-bath couplings. Floquet theory and time-dependent perturbation methods are used to facilitate both analytical and numerical solutions. We develop a theoretical formalism that provides a powerful tool for the detailed analysis of the dissipative quantum dynamics of multilevel systems driven by intense stochastic laser fields. It is found that the system relaxes to a steady state by the effect of laser phase and frequency noises and the kinetic of this relaxation increases with the addition of dissipative terms, introduced by the coupling to the reservoir. The case of amplitude fluctuations shows a different behavior. Other results concerning the destruction of quantum coherence and the dynamical localization will be established and further relaxation mechanisms such spontaneous emission and ionization process will be also considered.

# I- INTRODUCTION

The study of the time evolution of quantum systems, which are on the one hand driven by an external field and in contact with a heat bath (reservoir) on the other hand, has received a great deal of attention in recent years [1-4]. In quantum optics, such systems are investigated in the dressed-atom picture of resonance fluorescence [5], where a beam of atoms interacts with a coherent laser field and all the electromagnetic modes of the vacuum [6]. Moreover, it is by now recognized that nearly all types of laser–atom interactions can be strongly affected by laser noise. Indeed one practical reason to this fact is the use, in experiments, of high powers which are obtained in pulsed operation, and thus at the expense of poorly stabilized laser beams. Furthermore, real atoms experience a fluctuating environment of many perturbing interactions and ideal lasers exist only in theoretical models, while the used laser sources are subjected to many types of fluctuations notably in phase, amplitude and frequency [7-10]. Other kinds of fluctuations due to collisional effects can affect the atomic transition frequencies [9,11]. Therefore, we cannot establish, without taking into account of the dissipative action of the environment and the statistical properties of the laser light, a rigorous comparison between theoretical predictions and experimental results.

Different approaches to the dissipative dynamics of open quantum systems in strong external fields has been proposed and applied to the description of atoms under the influence of thermal noise [1-4]. While for the incoherence of laser field, a series of models, all based on
so-called pre-Gaussian Markovian processes [7-10] has been used in order to describe the stochastic behavior of the laser field. It is important to mention a few technical features of this model. It is based on the two-state random telegraph. It is not a Gaussian model but rather

"pre-Gaussian" and has a Gaussian limit [8]. Our choice of the random telegraph is based on the simplicity of this model that permits a unified treatment of different noisy laser in phase, amplitude and frequency. Several works have reported on the action of random process on a two–level system [7-10,12-15], particularly the evolution populations $\sigma_{nn}$ and the ionization probability.

In the present paper, we want to elucidate the role of quantum dissipation and laser noise on the atomic response. For this purpose, we derive a master equation, which provides a general framework for the dynamics of atoms interacting with strong laser noisy and thermal reservoir.

The basic idea underlying the theoretical formalism of that paper is to take into account the exact dynamics of the interaction between atomic system and external field by employing the Floquet basis for the reduced system rather than the stationary unperturbed states [1,2]. The interaction of the system (laser–atom) with the reservoir will be treated by the time-dependent perturbation theory, this treatment leads to a generalized quantum master equation for the reduced density matrix. Such statistical fundamental equation firstly introduced in quantum optic by Burshtein [16-18], contains all information concerning, respectively, the atomic transition dynamics, the stochastic evolution of laser field fluctuations and the dissipative mechanisms. We are concerned here with an important theme of contemporary research, namely the interplay between quantum coherence and external noise. In fact, the destruction of quantum coherence by noise is central to many fields and is reflected in the large number of papers recently published on this subject [19-23].

Since our computations are made at exact resonance, where the effects of spontaneous emission are important [24] and for strong laser field, where the probability to realize an ionization of atoms is highest. Therefore we shall here extend our theory by the inclusion of the relaxation rates corresponding to the spontaneous emission and the ionization processes and present the corresponding numerical results.

The theory is developed in section 2, by considering the case of strong laser–atom interaction in the presence of laser noise and dissipative effects, which are introduced by the coupling to the reservoir. Within the framework of the Floquet representation and the Markovian approximation, detailed theoretical calculations are feasible to obtain the reduced matrix density elements. The account of Floquet theory given here is rather brief, since the theory has been discussed at length in the recent literature (see e. g. refs. [25-29]). Moreover, the influence of ionization process on the response atomic function is presented. Numerical results concerning a model of two–level system is presented in section 3. At the end a summary of our results is given.

## II- THEORY

We consider in this paper an atomic system, which interacts with an external classical laser field. Moreover, the system (laser-atom) is coupled to a quantified radiation field in thermal equilibrium. In the following we will consider the behavior of the atom coupled to a reservoir with many degrees of freedom.

The aim of this paper is to provide a description of the dynamics in terms of the degrees of freedom of the atomic system alone by elimination of the reservoir variables. Since the atoms are driven strongly by an external laser field, our master equation is based on the atomic Floquet states rather than the unperturbed atomic states.

The total Hamiltonian governing the dynamics of the coupled system of matter and radiation degrees of freedom takes the form

$$H(t) = H_{A-L}(t) + H_I + H_R \tag{1}$$

where $H_{A-L}(t)$ is the total Hamiltonian of the atomic system and the external laser field, without interaction with the reservoir, given by

$$H_{A-L}(t) = H_A + V(t) + H_{SE} \tag{2}$$

with $H_A$ denotes the stationary atomic Hamiltonian, $V(t)$ the dipole interaction between the atomic system and the laser field and $H_{SE}$ the Hamiltonian of the simultaneous emission, which reads

$$H_{SE} = -\frac{\hbar}{2}\Gamma \tag{3}$$

where $\Gamma$ is a diagonal matrix composed by the Einstein coefficients of spontaneous emission process and defined by [29]

$$\Gamma_{nn} = \sum_{n'\langle n} \gamma_{nn'}^{SE} \qquad (4)$$

here $\gamma_{nn'}^{SE}$ are the radiative decay rates.

The Hamiltonian that describes the coupling between the matter degrees of freedom and the quantified radiation field may be written in the dipole approximation as

$$H_I = \hbar \sum_j z \left( \gamma_j a_j + a_j^+ \gamma_j^* \right) \qquad (5)$$

where $\gamma_j$ are the coupling constants. $a_j^+$ and $a_j$ are the quantum operators of creation and annihilation, z denotes the component of the dipole operator on the (**OZ**) axis. The free Hamiltonian of the reservoir is represented by

$$H_R = \hbar \sum_j \omega_j \left( a_j^+ a_j + \frac{1}{2} \right) \qquad (6)$$

with $\omega_j$ is a frequency corresponding to the $j^{eme}$ mode of the free quantified radiation.

The Von Neumann equation for the statistical operator $\rho$ of the total system reads

$$i\hbar \frac{\partial \rho}{\partial t} = [H(t), \rho] \qquad (7)$$

We introduce the interaction representation for treating the equation (7), and we put

$$H_0(t) = H_{A-L}(t) + H_R \qquad (8)$$

that is considered as the time-dependent unperturbed Hamiltonian, the evolution operator corresponding to this Hamiltonian is given by

$$U(t) = U_0(t) \otimes U_R(t) \qquad (9)$$

and

$$U(t) = \left[\exp\left\{-\frac{i}{\hbar}\int_0^t H_{A-L}(t')\,dt'\right\}\right]_+ \exp\left(-\frac{i}{\hbar}H_R t\right) \qquad (10)$$

where $[\ldots]_+$ is an operator of time ordering. In the interaction representation, the total density operator $\rho(t)$ and the interaction Hamiltonian $H_I$ take the following form

$$\tilde{\rho}(t) = U^+(t)\rho(t)U(t) \qquad (11)$$

and

$$\tilde{H}_I(t) = U^+(t)H_I U(t) \qquad (12)$$

and the dynamic equation (7) becomes

$$i\hbar \frac{\partial \tilde{\rho}(t)}{\partial t} = \left[\tilde{H}_I(t), \tilde{\rho}(t)\right] \qquad (13)$$

We assume that the interaction between the atomic system and the reservoir is weak that the coupling constants $\gamma_j \to 0$ and $\gamma_j^2 t = \text{Cst}$ for $t \to \infty$. In these conditions, the equation (13) will be treated by the time-dependent perturbation theory. At the second order in $\tilde{H}_I$, this last reads

$$\frac{\partial \tilde{\rho}(t)}{\partial t} = \frac{1}{i\hbar}\left[\tilde{H}_I(t), \tilde{\rho}(t_0)\right] - \frac{1}{\hbar^2}\int_{t_0}^{t} dt' \left[\tilde{H}_I(t), \left[\tilde{H}_I(t'), \tilde{\rho}(t')\right]\right] \tag{14}$$

In writing Eq. (14), it has been assumed that the interaction is adiabatically switched on at time ($t_0 \to -\infty$). Prior to this, the atomic system and the reservoir are uncorrelated and the total density matrix is given by the direct product

$$\tilde{\rho}(t_0) = \tilde{\sigma}(t_0) \otimes \rho_R \tag{15}$$

where $\tilde{\sigma}(t_0)$ is the reduced system density operator at initial time in the interaction representation and defined by the trace over the reservoir states. $\rho_R$ is the reservoir distribution function at equilibrium given by

$$\rho_R = \frac{\exp(-H_R / K_B T_R)}{Z_R} \tag{16}$$

here $Z_R$, $T_R$ are respectively the partition function and the temperature of the reservoir and $K_B$ is the Boltzman constant.

We need to define the Hamiltonian of interaction between the atomic system and the laser field without its coupling with the bath. In the dipole approximation, it reads

$$V(t) = e\, F_0\, z \cos(\omega t + \varphi(t) + x(t)) \tag{17}$$

in the case of phase fluctuations,

or
$$V(t) = e\, F_0\, (1 + x(t))\, z \cos(\omega t + \varphi(t)) \tag{18}$$

in the case of amplitude fluctuations, where $\omega$ is the laser frequency and e is the electron charge. $F_0$ is the electric field amplitude (possibly fluctuating in magnitude) and $\varphi(t)$ is the instantaneous phase of the laser (fluctuating around the mean value). In this paper, we use an intense laser field affected by a temporal stochastic process of jumps. These fluctuation mechanisms are described by the pre-Gaussian Markovian models [7-10]. In particular, we adopt the simplest example of two-state random telegraph, which is defined by $x(t) = \pm a$, where a is the amount of the jump assigned to the stochastic signal.

Since the telegraph process that we are considering here is Markovian, the conditioned probability density function associated with it, namely $p(s, t \mid s_0, t_0)$, is shown to satisfy the following Chapman–Kolmogorov equation [7-10,30]

$$\frac{\partial}{\partial t} p(s, t \mid s_0, t_0) = -\frac{1}{T} p(s, t \mid s_0, t_0) + \frac{1}{T} p(-s, t \mid s_0, t_0) \tag{19}$$

here $s_0$ is the initial state of random telegraph at the time $t_0$.

In the compact form [31], the equation (19) write as

$$\frac{d\mathbf{P}_s}{dt} = \sum_{s'} \mathbf{W}_s^{s'} \mathbf{P}_{s'} \tag{20}$$

where $\mathbf{W}_s^{s'} = \frac{1}{T}\begin{bmatrix} -1 & 1 \\ 1 & -1 \end{bmatrix}$ is the relaxation matrix composed by the frequencies of telegraph jumps process, where s and s' are two different states of random telegraph (s = 1,2) and corresponding to the telegraph signal amplitude $\{-a, +a\}$. T denotes the dwell time (i.e., the mean time between interruptions) of the telegraph. In the following, and in presence of the noise, all physical operators, such reduced density operator, interaction Hamiltonian, dipole operator…, will be indiced by s in indication of the stochasticity influence.

By elimination of the reservoir variables in Eq. (14), we have

$$\dot{\tilde{\sigma}}_s(t) = -\frac{1}{\hbar^2} \int_{t_0}^{t} dt' \, \mathrm{Tr}_R \left[\tilde{H}_{Is}(t), \left[\tilde{H}_{Is}(t'), \tilde{\sigma}_s(t')\right]\right] \tag{21}$$

After tracing on the reservoir variables, we combine the Chapman-Kolmogorov equation (20), which represents the stochastic evolution of the random telegraph, to the equation (21) representing the atomic dynamics. A master equation for the reduced density operator is derived in the interaction picture

$$\dot{\tilde{\sigma}}_s(t) = \sum_{s'} \mathbf{W}_s^{s'} \tilde{\sigma}_{s'}(t) + \tilde{D}_s(t) \tag{22}$$

where

$$\tilde{D}_s(t) = -\frac{1}{\hbar^2} \int_{t_0}^{t} dt' \, \mathrm{Tr}_R \left[\tilde{H}_{Is}(t), \left[\tilde{H}_{Is}(t'), \tilde{\sigma}_s(t')\right]\right] \tag{23}$$

is the time-dependent operator describing the dissipative effects induced by the coupling to the reservoir. In writing the expression of $\tilde{D}_s(t)$, we have replace up $\tilde{\rho}_s(t) \approx \tilde{\sigma}_s(t) \otimes \rho_R$ to second order in the coupling constants by its zeroth-order approximate. Since $\tilde{H}_{Is}(t)$ is a periodic function in time, we explicitly construct the operator $U_0(t)$. To this end, it is necessary to treat the interaction with the strong incoherent laser field exactly, and solve the corresponding Schrodinguer equation generated by the Hamiltonian $H_{A-L}(t)$ by using the Floquet theory. According to Floquet's theorem [25-29], there exists a complete set of solutions labeled by quantum numbers $\alpha$ of the form

$$\left|\psi_s^\alpha(t)\right\rangle = \exp(-i\varepsilon_\alpha t/\hbar)\left|\phi_{\alpha,s}(t)\right\rangle \tag{24}$$

Where $\hbar\varepsilon_\alpha$ and $\left|\phi_{\alpha,s}(t)\right\rangle$ are respectively the quasi-energies and the eigenstates of Floquet. The time-evolution operator $U_0(t)$ for the matter degrees of freedom in the Floquet representation [1,2], is given by

$$U_0(t) = \exp(i\varepsilon_\alpha t/\hbar)\left|\phi_{\alpha,s}(t)\right\rangle\left\langle\phi_{\alpha,s}(0)\right| \tag{25}$$

In the interaction picture, the operators annihilation, creation and dipole respectively $a_j$, $a_j^+$ and z take the form

$$\tilde{a}_j = U^+(t)a_j U(t) = \exp(-i\omega_j t)\, a_j, \tag{26}$$

$$\tilde{a}_j^+ = \exp(i\omega_j t)a_j^+ \tag{27}$$

and

$$\tilde{z}_s(t) = U^+(t)zU(t) = \sum_{\alpha\beta k}' \exp(i\varepsilon_{\alpha\beta}(k)t)Z_{\alpha\beta,s}(k)|\phi_{\alpha,s}(t)\rangle\langle\phi_{\alpha,s}(0)| \qquad (28)$$

where

$$Z_{\alpha\beta,s}(k) = \frac{\omega}{2\pi}\int_0^{\frac{2\pi}{\omega}} dt\, \exp[-ik\omega t]\langle\phi_{\alpha,s}(t)|z|\phi_{\beta,s}(t)\rangle \qquad (29)$$

are the dipole matrix elements between Floquet states. The symbol 'prime' in the sum Eq. (28) indicates that only the triplets $(\alpha,\beta,k)$ that verify the condition $\varepsilon_\alpha - \varepsilon_\beta + k\hbar\omega > 0$ will be considered, in the purpose to eliminate the degenerate Floquet eigenstates. In order to calculate the dissipation operator $\tilde{D}_s(t)$, we follow the methodology formulated in the Ref. [1], indeed, we have

$$\tilde{H}_{Is}(t) = \sum_{\alpha\beta k}' \sum_j \gamma_j \left[e^{i(\varepsilon_{\alpha\beta}(k)-\omega_j)t}F_{\alpha\beta,s}\,a_j + H.c.\right] \qquad (30)$$

with H.c. is the hermitic conjugate,

$$F_{\alpha\beta,s} = |\phi_{\alpha,s}(0)\rangle\langle\phi_{\beta,s}(0)| \qquad (31)$$

and

$$\varepsilon_{\alpha\beta}(k) = \varepsilon_\alpha - \varepsilon_\beta + k\hbar\omega \qquad (32)$$

where $-\infty < k < +\infty$. By taking into account of the (Eq. 30), $\tilde{D}_s(t)$ reads in the form

$$\tilde{D}_s(t) = \sum_{\alpha\beta k}{}' \sum_{\alpha'\beta'k'}{}' \sum_j \gamma_j^2 \int_{t_0}^{t} dt' \Big\{ \langle a_j^+ a_j \rangle \big( A(t,t') F_{\alpha\beta,s} \tilde{\sigma}_s(t') F_{\beta'\alpha,s} - A^+(t,t') F_{\beta\alpha,s} F_{\alpha'\beta',s} \tilde{\sigma}_s(t') \big)$$
$$+ \langle a_j a_j^+ \rangle \big( A^+(t,t') F_{\beta\alpha,s} \tilde{\sigma}_s(t') F_{\alpha\beta',s} - A(t,t') F_{\alpha\beta,s} F_{\beta'\alpha,s} \tilde{\sigma}_s(t') \big) + H.c. \Big\} \quad (33)$$

with

$$A(t,t') = Z_{\alpha\beta,s}(k) Z^*_{\beta'\alpha',s}(k') \exp\big(i(\varepsilon_{\alpha\beta}(k) - \omega_j)t\big) \exp\big(i(\varepsilon_{\alpha'\beta'}(k') - \omega_j)t'\big) \quad (34)$$

and

$$\langle a_j^+ a_j \rangle = N(\omega_j) = \left( \exp\left( \frac{\hbar\omega_j}{K_B T_R} \right) - 1 \right)^{-1} \quad (35)$$

$N(\omega_j)$ is the photon number operator. Since the freedom degrees of the bath are infinite, we can make the substitution $\sum_j \gamma_j^2 ... \rightarrow \int d\omega J(\omega)...$, where $J(\omega)$ is a function which is proportional to the bath spectral density. In order to perform the integration, which is present in the expression of $\tilde{D}_s(t)$ Eq. (33), further conditions must be imposed on the reservoir in the purpose to prevent the energy, initially in the atomic system, from returning back from the heat bath to the system in any finite time, i.e., (treat the coupling of the reduced system to the reservoir as an irreversible process). At this stage, we make two approximations.

i) Equation (33) contains $\tilde{\sigma}_s(t')$ in the integral, and hence the behavior of the atomic system depends on its past history from $t' = t_0$ to $t' = t$. The motion of the atomic system is however, damped by the coupling to the reservoir and damping destroys the knowledge of the past behavior of the system. Therefore the first assumption is that $\dot{\tilde{\sigma}}(t)$ depends only on it's present value $\tilde{\sigma}_s(t)$ (Markovian approximation) [1,2].

ii) Let us consider an operator **B** of the bath and it's time correlation function $\langle \mathbf{B}(t-t')\mathbf{B}^+ \rangle$. Since the reservoir is assumed to be large and Markovian. Thus it is expected that $\langle \mathbf{B}(t-t')\mathbf{B}^+ \rangle$ will be nonzero for some time interval $t - t' < \tau_R$, where $\tau_R$ is the correlation time of the reservoir. Interactions at times t and t' become progressively less correlated for $t - t' \gg \tau_R$. The correlation function $\langle \mathbf{B}(t-t')\mathbf{B}^+ \rangle$ is only maximum at $t = t'$. We can therefore tend the superior born of integration in Eq. (33) to the infinite ($t \to \infty$).

With these two approximations and by using the following expression for the initial time ($t_0 \to -\infty$),

$$\int_{-\infty}^{+\infty} dt' \exp\left(i\left(\omega - \varepsilon_{\alpha'\beta'}(k')\right)t'\right) = 2\pi \delta\left(\omega - \varepsilon_{\alpha'\beta'}(k')\right) \tag{36}$$

The integro-differential equation reads

$$\tilde{D}_s(t) = 2\pi \sum_{\alpha\beta k} \sum_{\alpha'\beta'k'}{}' J(\varepsilon_{\alpha'\beta'}(k'))\{N(\varepsilon_{\alpha'\beta'}(k'))(A'(t)F_{\alpha\beta,s}\tilde{\sigma}_s(t')F_{\beta\alpha,s} - A'^+(t)F_{\beta\alpha,s}F_{\alpha'\beta',s}\tilde{\sigma}_s(t))$$
$$+ (1 + N(\varepsilon_{\alpha'\beta'}(k')))(A'^+(t)F_{\beta\alpha,s}\tilde{\sigma}_s(t)F_{\alpha'\beta',s} - A'(t)F_{\alpha\beta,s}F_{\beta'\alpha',s}\tilde{\sigma}_s(t)) + \text{H.c.}\} \tag{37}$$

Where

$$A'(t) = \exp\left(i\left(\varepsilon_{\alpha\beta}(k) - \varepsilon_{\alpha'\beta'}(k)\right)t\right) Z_{\alpha\beta,s}(k) Z^*_{\alpha'\beta',s}(k'). \tag{38}$$

This last quantity is maximal for

$$\varepsilon_{\alpha\beta}(k) - \varepsilon_{\alpha'\beta'}(k') = 2n\pi \tag{39}$$

Where n is a positive or negative integer, for the case of $n = 0$ only terms such as $(\alpha, \beta, k) = (\alpha', \beta', k')$ must be kept in Eq. (37). The equation (33) takes the final form

$$\tilde{D}_s(t) = \sum_{\alpha\beta k}{}' \Omega_{\alpha\beta}(k) \left\{ N(\varepsilon_{\alpha\beta}(k)) \left( \left[ F_{\alpha\beta,s} \tilde{\sigma}_s(t), F_{\beta\alpha,s} \right] + \left[ F_{\alpha\beta,s}, \tilde{\sigma}_s(t) F_{\beta\alpha,s} \right] \right) \right. \\ \left. + \left(1 + N(\varepsilon_{\alpha\beta}(k))\right) \left( \left[ F_{\beta\alpha,s} \tilde{\sigma}_s(t), F_{\alpha\beta,s} \right] + \left[ F_{\beta\alpha,s}, \tilde{\sigma}_s(t) F_{\alpha\beta,s} \right] \right) \right\} \quad (40)$$

with

$$\Omega_{\alpha\beta}(k) = 2\pi J\big(\varepsilon_{\alpha\beta}(k)\big) \big| Z_{\alpha\beta,s}(k) \big|^2 \quad (41)$$

By projecting on the Floquet basis $\{|\phi_{\alpha,s}(0)\rangle\}$, the master equation for the diagonal and the off-diagonal elements respectively $\tilde{\sigma}_{\alpha\alpha,s}(t)$ and $\tilde{\sigma}_{\alpha\beta,s}(t)$ read

$$\dot{\tilde{\sigma}}_{\alpha\alpha,s}(t) = \sum_{s'} W_s^{s'} \tilde{\sigma}_{\alpha\alpha,s'}(t) + \sum_{\gamma}{}' \left( M_{\gamma\alpha} \tilde{\sigma}_{\gamma\gamma,s}(t) - M_{\alpha\gamma} \tilde{\sigma}_{\alpha\alpha,s}(t) \right) \quad (42)$$

and

$$\dot{\tilde{\sigma}}_{\alpha\beta,s}(t) = \sum_{s'} W_s^{s'} \tilde{\sigma}_{\alpha\beta,s'}(t) - \frac{1}{2} \left[ \sum_{\gamma}{}' \left( M_{\alpha\gamma} + M_{\beta\gamma} \right) \right] \tilde{\sigma}_{\alpha\beta,s}(t) \quad (43)$$

where the coefficients $M_{\alpha\beta}$ are defined by

$$M_{\alpha\beta} = 2 \sum_k \left\{ \left(1 + N\big(\varepsilon_{\alpha\beta}(k)\big)\right) \Omega_{\alpha\beta}(k) + N\big(\varepsilon_{\alpha\beta}(k)\big) \Omega_{\beta\alpha}(k) \right\} \quad (44)$$

and their solutions are given by

$$\tilde{\sigma}_{\alpha\alpha,s}(t) = \sum_{\beta,s'} [\exp(-\Lambda 1 t)]_{\alpha\beta,ss'} \tilde{\sigma}_{\beta\beta,s'}(0) \qquad (45)$$

and

$$\tilde{\sigma}_{\alpha\beta,s}(t) = \sum_{\gamma,s'} [\exp(-\Lambda 2 t)]_{\alpha\gamma,ss'} \tilde{\sigma}_{\gamma\beta,s'}(0) \qquad (46)$$

where

$$\Lambda 1_{\alpha\beta,ss'} = -W_s^{s'} \delta_{\alpha\beta} - \left( M_{\beta\alpha} - \delta_{\alpha\beta} \sum_{\eta} M_{\alpha\eta} \right) \delta_{ss'} \qquad (47)$$

and

$$\Lambda 2_{\alpha\gamma,ss'} = -W_s^{s'} \delta_{\alpha\gamma} + \frac{1}{2} \delta_{\alpha\gamma} \delta_{ss'} \left[ \sum_{\eta}' \left( M_{\alpha\eta} + M_{\beta\eta} \right) \right] \qquad (48)$$

The theoretical expressions for populations and coherence of quasienergie states respectively Eq. (45) and Eq. (46) have to be transformed back into the atomic basis which yields $\sigma_{nn}$ and $\sigma_{nn'}$. In Schrödinguer picture we then obtain

$$\dot{\sigma}_s(t) = -\frac{i}{\hbar}[H_0(t), \sigma_s(t)] + \sum_{s'} W_s^{s'} \sigma_{s'}(t) + D_s(t) \qquad (49)$$

with

$$D_s(t) = U_0 \tilde{D}_s U_0^+ \qquad (50)$$

$$D_s(t) = \sum_{\alpha\beta k}' \Omega_{\alpha\beta}(k) \{ (1 + N(\varepsilon_{\alpha\beta}(k))) \left( [R^+_{\alpha\beta,s}(t), \sigma_s(t) R_{\alpha\beta,s}(t)] + [R^+_{\alpha\beta,s}(t) \sigma_s(t), R_{\alpha\beta,s}(t)] \right)$$
$$+ N(\varepsilon_{\alpha\beta}(k)) \left( [R_{\alpha\beta,s}(t), \sigma_s(t) R^+_{\alpha\beta,s}(t)] + [R_{\alpha\beta,s}(t) \sigma_s(t), R^+_{\alpha\beta,s}(t)] \right) \} \qquad (51)$$

where

$$R_{\alpha\beta,s}(t) = U_0(t) F_{\alpha\beta,s} U_0^+(t) = \exp\left(-i(\varepsilon_\alpha - \varepsilon_\beta)t\right) |\phi_{\alpha,s}(t)\rangle\langle\phi_{\beta,s}(t)| \qquad (52)$$

The main difficulty of typical problems lies in the correct averaging of the matrix density over all realizations of noise. In fact, what is physically wanted is $\langle\sigma_{nn'}\rangle$, that is, the solution to the master equation in the atomic states and averaged over the ensemble of jumps of the implicit telegraph x (t). To obtain $\langle\sigma_{nn'}\rangle$ one proceeds indirectly, by defining a marginal average $\sigma_{nn',s}(t)$, given by the equation

$$\langle\sigma_{nn'}\rangle = \sum_s g(s)\, \sigma_{nn',s} \qquad (53)$$

where g(s) is the initial probability distribution of the random process and $\sigma_{nn',s}(t)$ the average value of $\sigma_{nn'}(t)$ under the condition that x (t) is fixed at the value s at time t. By projecting on the atomic basis $\{|n\rangle\}$, the master equation in the Schrödinguer picture then finally reads

$$\dot{\sigma}_{mn,s}(t) = \sum_{s'} W_s^{s'} \sigma_{mn,s'}(t) + \sum_{m'} \left(D1_{m'n,s}(t) + \frac{i}{\hbar} H_0(t)_{m'n,s}\right) \sigma_{mm',s}(t) +$$
$$\sum_{m'} \left(D1_{mm',s}(t) - \frac{i}{\hbar} H_0(t)_{mm',s}\right) \sigma_{mm',s}(t) + \sum_{m'n'} D2_{mm'n'n,s}(t) \sigma_{m'n',s}(t) \qquad (54)$$

with the two terms responsible of the dissipation are defined by

$$D1_{mn,s}(t) = -\sum_{\alpha\beta k}{}' \Omega_{\alpha\beta}(k) \left\{ \left(1 + N(\varepsilon_{\alpha\beta}(k))\right) \phi_{m\alpha,s}(t)\, \phi^{+}_{n\alpha,s}(t) + \right.$$
$$\left. N(\varepsilon_{\alpha\beta}(k))\, \phi_{m\beta,s}(t)\, \phi^{+}_{n\beta,s}(t) \right\} \quad (55)$$

and

$$D2_{mm'n'n,s}(t) = 2\sum_{\alpha\beta k}{}' \Omega_{\alpha\beta}(k) \left\{ \left(1 + N(\varepsilon_{\alpha\beta}(k))\right) \phi_{m\beta,s}(t)\, \phi^{+}_{m'\alpha,s}(t)\, \phi_{n'\alpha,s}(t)\, \phi^{+}_{n\beta,s}(t) + \right.$$
$$\left. N(\varepsilon_{\alpha\beta}(k))\, \phi_{m\alpha,s}(t)\, \phi^{+}_{m'\beta,s}(t)\, \phi_{n'\beta,s}(t)\, \phi^{+}_{n\alpha,s}(t) \right\} \quad (56)$$

where $\phi_{n\alpha,s}(t)$ are the Floquet states, which are projected on the atomic basis $\{|n\rangle\}$.

It is important to note that the general master equation (54) contains dissipative terms Eq. (55) and Eq. (56) that explicitly depend on time. This is the main difference to the usual optical Bloch equations. The physical interpretation to this fact is the strong distortion of the atomic dipole moment, which is induced by the external laser field. Since the atom couples to the environment via its dipole moment, the laser field also strongly influences the dissipation process [2].

## III- Results and discussion.

In this section we gather typical numerical results for the excitation and ionization of two-level atoms by strong laser fields in the presence of noise and dissipation mechanisms. To illustrate the effects of dissipation and laser noise on the atomic response, we present the evolution of atomic populations $\sigma_{nn}(t)$, which are obtained by numerical integration of the master equation (54). Our theoretical formalism is valid for the general case of multilevel systems but in order to keep the discussion simple; we will restrict our application to the two-level atoms for which a detailed study of the dissipative non-linear dynamics will be presented. A particular attention will be paid to the case of strong laser field, where the dipole operator is taken between the Floquet eigenstates $\left\{\left|\phi_{\alpha,s}(t)\right\rangle\right\}$ rather than between unperturbed atomic states $\left\{|n\rangle\right\}$.

Having established the effects of strong laser noise on the atomic response and explored some features of different sources of noise (phase, amplitude and frequency). We concentrate our attention in this paper to the examination of quantum dissipation induced by coupling to the reservoir and when the noise is added to the laser field. We choose the inverse Rabi frequency $\Omega$ as time unit in the aim to analyze the obtained results in term of the noise magnitude. We are interested by a large light intensity such that the Rabi frequency is set to be equal to the atomic unit ($\Omega$ = 1 a.u.), This certainly is a very strong intensity.

## III. 1 Dissipative nonlinear quantum dynamics in the excitation of two-level systems.

We begin by representing only the effects of quantum dissipation on the atomic response. Figures 1(a) and 1(b) show time evolution of populations of a two-level atomic system driven by strong *coherent laser field* sufficiently intense to remove a significant fraction of the population from the atomic ground state. One might think that the only consequence of a field this intense would be to lower the overall of the atom. The optical field is nearly resonant with allowed transition between discrete states of two-level atom. In absence of spontaneous emission decay and ionization effects, Figure 1(a) represents the atomic response without dissipation, the atomic system oscillates between the ground state $|1\rangle$ and some other discrete level $|2\rangle$ and we have the ordinary Rabi oscillations. In Figure 1(b) the dissipation effect introduces a damping of Rabi oscillations. If damping effects are present, we expect that the Rabi oscillations will eventually become damped out and that the population's difference will approach some steady-state value for large scale of time. Hence, Rabi oscillations are not present in the steady state.

It's interesting to note the presence of an irregular behavior on the oscillations of the two populations for a strong laser field; in fact we observe small oscillations which come to superpose to the Rabi oscillations, their amplitude is weak and disappears when the electric field strength $F_0$ becomes small with respect to the atomic unit of field strength. These little oscillations represent the fast no-rotating variable phases $\exp(\pm i(\omega+\omega_{nn'})t)$ in laser-atom interaction, which is treated in a non-perturbative way (Floquet theory).

Figure 2, shows some typical results, in fact we take into account of both laser phase noise and dissipation influences. Taking a phase jump $a = 0.4\pi$ and three switching rates ($\Omega T = 0.1$, 1 and 10). We display the time evolution populations on two column, in the column (A), only the effect of laser noise is considered, we remark in this case that for a switching rate ($\Omega T = 1$), i.e., the noise frequency ($1/T$) is of the same size order that the Rabi frequency $\Omega$, a destruction of the atomic coherence is observed. The damping rate is strong and the relaxation to the steady state is rapid. The Rabi oscillations are restored when we consider the case of slow fluctuations ($\Omega T = 10$) and fast fluctuations ($\Omega T = 0.1$) and the damping rate is weak. The column (B) represents the same situation but by introducing moreover of the laser noise the dissipative terms (see Eqs (55) and (56)). A similar behavior is remarked as the column (A), but with damping rate more intense. For a switching rate ($\Omega T = 0.1$), a partial destruction of the atomic coherence is induced by the dissipation effects. The kinetic of populations relaxation is more rapid that in the column (A). We remark that one of the effects of the quantum dissipation is the breaking of the atomic coherence especially for the case of ($\Omega T = 0.1$) and the establishment of the dynamical localization regime for the case of ($\Omega T = 1$). In Figure 3, we plot the time evolution of two-level atom populations in two situations, in first time, by neglecting the effect of quantum dissipation (column (A)) and considering only an amplitude telegraph noise and in second time we combine the two influences of noise and quantum dissipation. Taking an amplitude jump $a = 0.1$ a.u. and three different switching rates ($\Omega T = 0.1$, 1 and 100). The column (A) shows pronounced quasi-oscillations, we remark a very weak damping at ($\Omega T = 1$ and $\Omega T = 100$) and rapid relaxation for ($\Omega T = 0.1$) with respect to the both cases of ($\Omega T = 1$ and 100). In order to lead the system to the steady state we must use a large number of Rabi periods rather than in the case of phase fluctuations. The column (B) shows closely similar behavior that in column (A). The complicated structure of these curves is a consequence of the action of amplitude laser noisy on the reduced system dynamics. In fact, one observes a separation between the two occupation probabilities, every population $\sigma_{11}(t)$ and $\sigma_{22}(t)$ performs independently

irregular oscillations, which converge to stationary state. In constraints to the case of phase, the addition of dissipation in Figure 3 (column (B)) introduces a weakness of the damping.

Figure 4 illustrates the case of frequency fluctuations, this kind of noise is introduced by collisional effects, indeed the transition frequency $\omega_{21}$ can also fluctuate around its fixed value. The simplest model of such interruption collisions [9,11] assumes that the atomic transition frequency $\omega_{21}$ should be replaced by $\omega_{21}(t) = \omega_{21} + x(t)$. By taking a jump parameter a = 0.1 a.u. and three different frequency switching rates ($\Omega T$ = 1, 10 and 100). We remark in the column (A) damped quasiperiodic oscillations. The case of ($\Omega T$ = 10) corresponds to strong damping without any convergence to a steady state. While the relaxation to an equilibrium state of value 1/2 is clear for a switching rate ($\Omega T$ = 1). The damping becomes weak for ($\Omega T$ = 100) and two independent beats phenomena are observed. The complicated time evolution of populations is a result of Rabi oscillation interference. In the column (B) where we take into account of quantum dissipation, the two populations relax to equilibrium state. The thermal noise induced by coupling to the bath introduces a complete destruction of the atomic coherence. The kinetic of relaxation and rate damping decrease from the case of fast fluctuations ($\Omega T$ = 0.1) to the slow fluctuations ($\Omega T$ = 100).

In Figures 2 and 4, the comparison between the two columns (A) and (B) shows that the dissipation which behaves as a noise (thermal noise) leads the system to an equilibrium state with rapid kinetic of relaxation. The damping rates become large when we introduce the dissipation terms (column B). In other hand, we clearly see the destruction of atomic coherence, which increases, when we take into account of dissipation . The dynamical localization regime appears for phase and frequency noises. However Figure 3 shows important asymmetries. This behavior is justified by the fact that in the case of amplitude fluctuations, the jump parameter a, assigned to stochastic process, appear in term of laser intensity $F_0$ (1 ± a ), while in the case of phase noisy , the dependence occurs in term of $\exp(\pm ia)$.

### III. 2  Dissipative nonlinear quantum dynamics in the ionization of two-level systems.

Having established a formal framework for the excitation of atoms by laser noisy in presence of the reservoir action and explored some of its general predictions. We turn now to the examination of ionization effects on the populations and the illustration of the modifications generated by the different kinds of noise and quantum dissipation on the ionization rates. In order to analyze the ionization effects we adopt the extended two-level system model proposed by Yeh and Eberly [14,32]. The computation of ionization probability is made by the incorporation of responsible term of ionization $[-\tau_{EC} \sigma_{mn,s}$, with $\tau_{EC} = R_{mc} \delta_{mn} + 1/2(R_{mc} + R_{nc})(1-\delta_{mn})$ and $R_{nc}$ is the relaxation rate from the excited state $|n\rangle$ to the continuum $|c\rangle$] in the motion equation (54). The trace of $\sigma$ over a complete set atomic states leads to the expression $P_{ion}(t) = 1 - \sum_{n=1}^{2} \sigma_{nn}$ for the total ionization probability of the system [4,13,33].

We begin by showing successively the effects of laser noise and reservoir dissipation on the ionization probability. As illustrated in Figure 5, we have plotted total ionization as a function of Rabi periods. Taking an intense electric laser field such as ($\Omega$ = 1 a.u.) and a resonant laser frequency. The four curves of Figure 5 correspond respectively to the situations where: noise and dissipation are neglected, only the noise is considered, only dissipation effect is retained and both noise and dissipation exist. In Figure 5(a) we have considered a phase noise of a = 0.4π, the results depend on the fluctuations time scale (1/T) compared to the other characteristic time scales of the problem such the Rabi frequency Ω. The minimum variations of the ionization probability are obtained, where we neglect both noise and dissipation. When we take into account of dissipation, the ionization probability increases. For large number of Rabi period, we remark that the phase fluctuations effect is more important that the reservoir action, in fact, the ionization probability rapidly increases when we introduce phase fluctuations corresponding to the case of (ΩT = 0.1). In presence of dissipation terms, the ionization probability variations remain closely constant. We conclude that noise and dissipation rapidly leads the atom to the ionization states. Figure 5(b) shows the case of amplitude noise, where the jump parameter is taken a = 0.1 a.u. and a switching rate (ΩT = 1). A close behavior is observed as Figure 5(a) but the amplitude noise effect is very weak. The technique that we have used will be applied subsequently to frequency noise.

A plot that gives a pictorial sense of how ionization proceeds in time is given in Figure 6. The total probability of ionization $P_{ion}(t)$ as a function of the Rabi period is plotted together with the occupation probabilities of bound states. The oscillations in these curves reflect the Rabi oscillations of the atom between the resonantly coupled states $|1\rangle$ and $|2\rangle$. These oscillations are damped by ionization in a few Ωt's. This behavior is well known from the study of bound states coupled by an intense field. The column (A) of Figure 6 shows the response of two–level in presence of phase noise and (B column) by considering moreover the effect of dissipation. The same parameters are taken as Figure 2. The populations, which have not been lost through direct ionization to the atomic continuum, oscillate in the same manner that in absence of ionization effect, but there is a progressive decay to a zero probability. The ionization probability can be viewed as dominant in a few Rabi periods and increases rapidly in time when the quantum dissipative effects are considered. Figure 7 displays the same behavior that Figure 6 but for an amplitude noise.

## CONCLUSION

In this paper we have investigated at length the non-linear dynamics of dissipative quantum atomic systems subjected to the action of heat bath and periodically driven by strong laser field which is affected by classical noise. We have derived and solved a master equation for atoms in strong noisy laser fields and in presence of reservoir dissipative effects. Such equation based on the Floquet states rather than the unperturbed atomic states has given typical and interesting results concerning the atomic dynamics. In fact, we have demonstrated how the master equation formalism, the Floquet theory, the pre-Gaussian models of laser noise and the Markovian coupling of quantum system to an environment can be combined together in order to tackle a general theoretical formalism and a powerful tool for the detailed analysis of the interaction of an atomic system with intense incoherent laser field and with a large reservoir.

We have shown that the dissipation terms, which are time dependent respect to those in the Bloch equations, force the system to settle to some '' preferred states '' it is the dynamical localization regime observed in the cases of phase and frequency noises, as we have explored in this paper. Moreover under the action of these decay mechanisms, the atomic system exhibits different regimes such as the destruction of coherence, the relaxation to equilibrium state. In general, the strength of damping and the kinetic of relaxation increase with the addition of dissipation effects but the case of amplitude fluctuations show a different behavior. We have also analyzed the modifications induced by ionization effects.

## Acknowledgments

It is a pleasure to thank the Professors A. Buchleitner, R. Graham and Heinz-Peter Breuer for sending us their interesting reprints concerning this subject. We also wish to thank Professors A. Maquet and C. J. Joachain for very helpful suggestions and communication.

# Figure captions

**Figure 1.** Populations $\sigma_{nn}$ versus time (in units of inverse Rabi frequency $\Omega$) for two–level atoms, resonantly excited by an intense electric laser field, such that the Rabi frequency is set $\Omega = 1$ a.u. The emission spontaneous coefficient is $\gamma_{21} = \Omega/10^5$ a.u.

(a) Both effects of noise and dissipation are neglected (coherent laser and no coupling to the reservoir).

(b) In absence of noise and the dissipation is considered.

**Figure 2.** Populations $\sigma_{nn}$ versus time (in units of inverse Rabi frequency $\Omega$) for two–level atom resonantly excited by random telegraph phase noise, successive frames are for different values of phase switching rates $\Omega T = 0.1$, 1 and 10. We use strong laser field, such that the Rabi frequency is set $\Omega = 1$ a.u. The phase jump parameter is $a = 0.4\pi$, the emission spontaneous coefficient is $\gamma_{21} = \Omega/10^5$ au.

 - The column (A) represents the effects of phase noise.

 - The column (B) represents the same situation as column (A) but by adding the dissipative effects.

**Figure 3.** Same as Figure 2., but for an amplitude noise with the jump parameter is $a = 0.1$ a.u. and three switching rates $\Omega T = 0.1$, 1 and 100.

**Figure 4.** Same as Figure 3., but for a frequency noise.

**Figure 5.** Ionisation probability $P_{ion}(t)$ versus time (in units of inverse Rabi frequency $\Omega$). Same parameters as the previous figures are used. The The relaxation rate from bound states to the continuum is $R_{2c} = \Omega/100$.

(a) Dotted line: $P_{ion}(t)$ in absense of noise and dissipation. Dashed line: only the effect of phase noise for $\Omega T = 0.1$. Dashed-Dotted line: only the effect of dissipation is considered and Solid line: both effects of noise and dissipation are considered.

(b) Same as (a), but for an amplitude noise with $\Omega T = 1$.

**Figure 6.** The columns A and B are, respectively, same as Figure 2(A) and Figure 2(B), but take into account of the ionisation process represented by the probability $P_{ion}(t)$. The relaxation rate from bound states to the continuum is $R_{2c} = \Omega/100$.

**Figure 7.** The columns A and B are, respectively, same as Figure 3(A) and Figure 3(B), but take into account of the ionisation process represented by the probability $P_{ion}(t)$. The relaxation rate from bound states to the continuum is $R_{2c} = \Omega/100$.

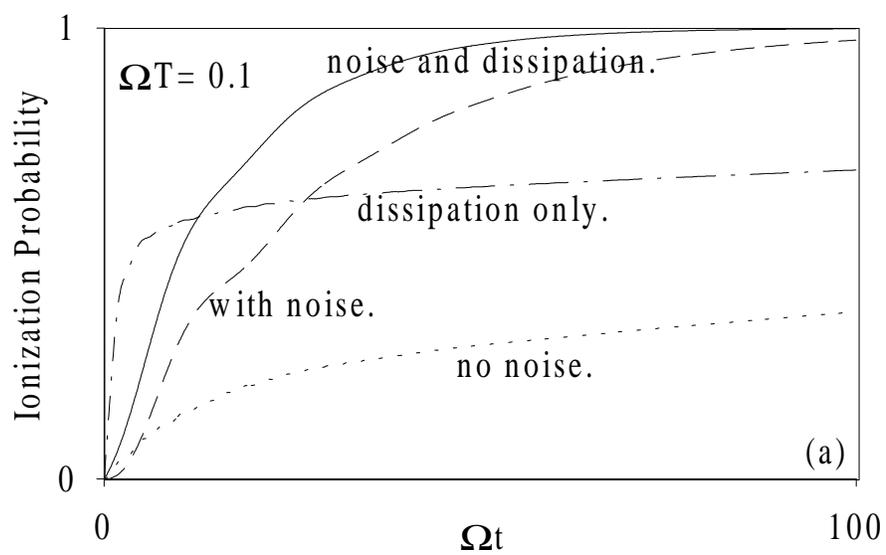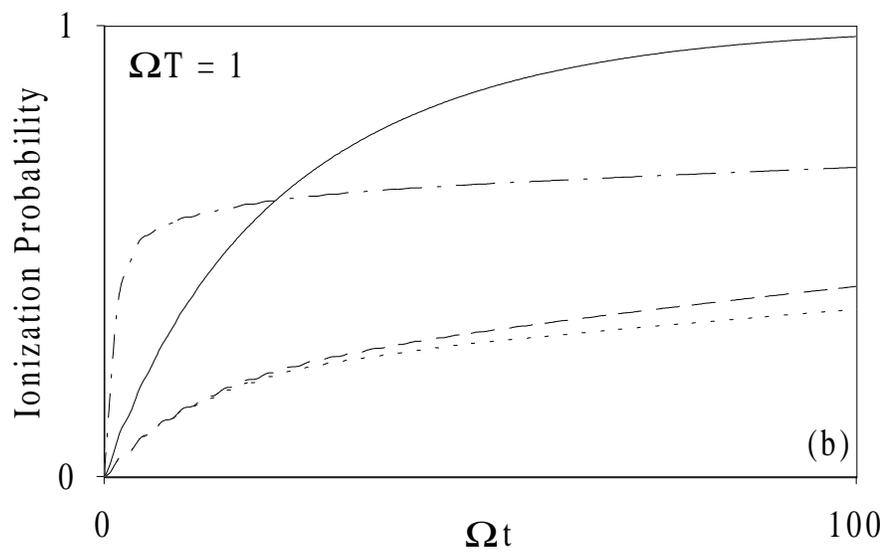

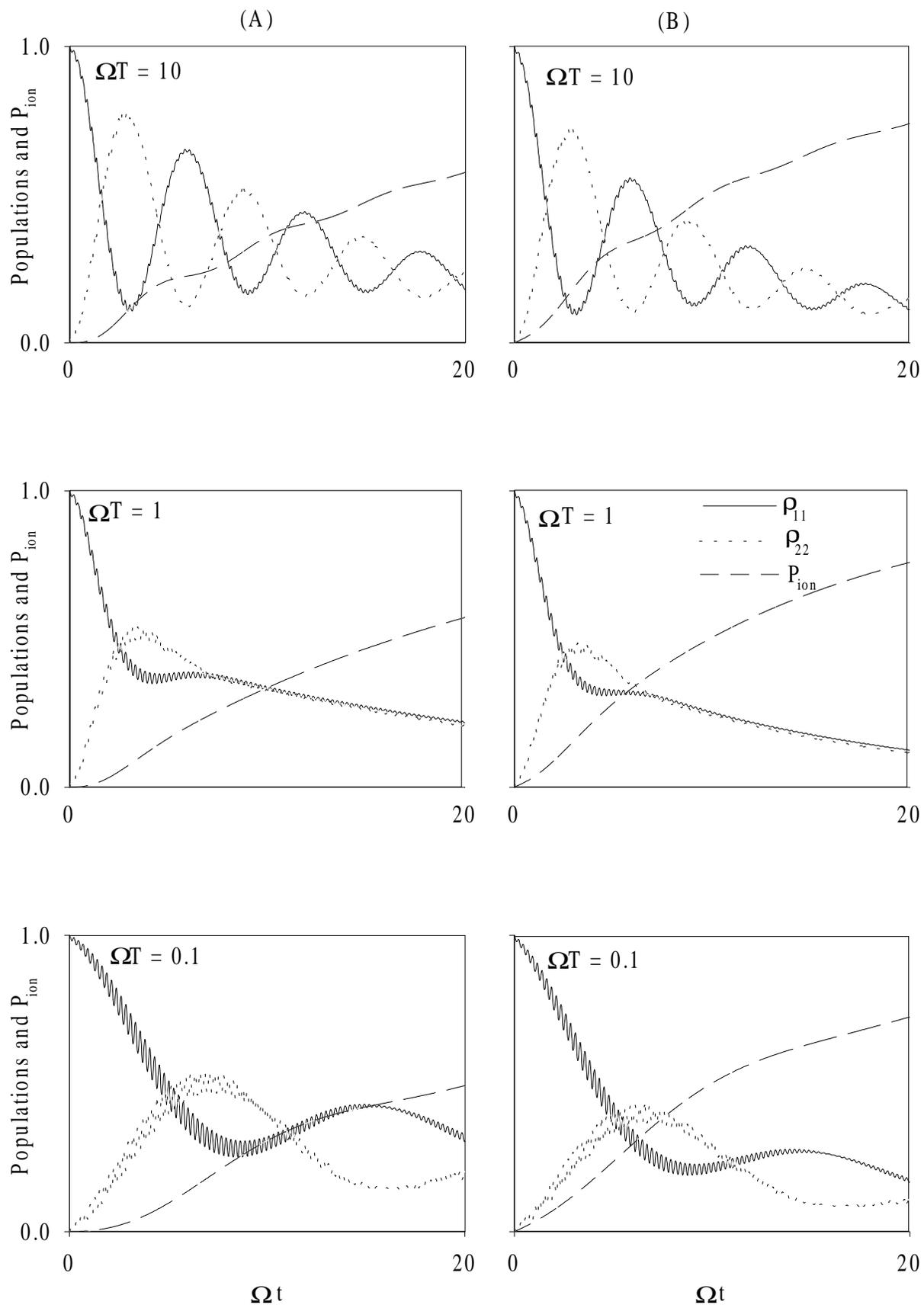

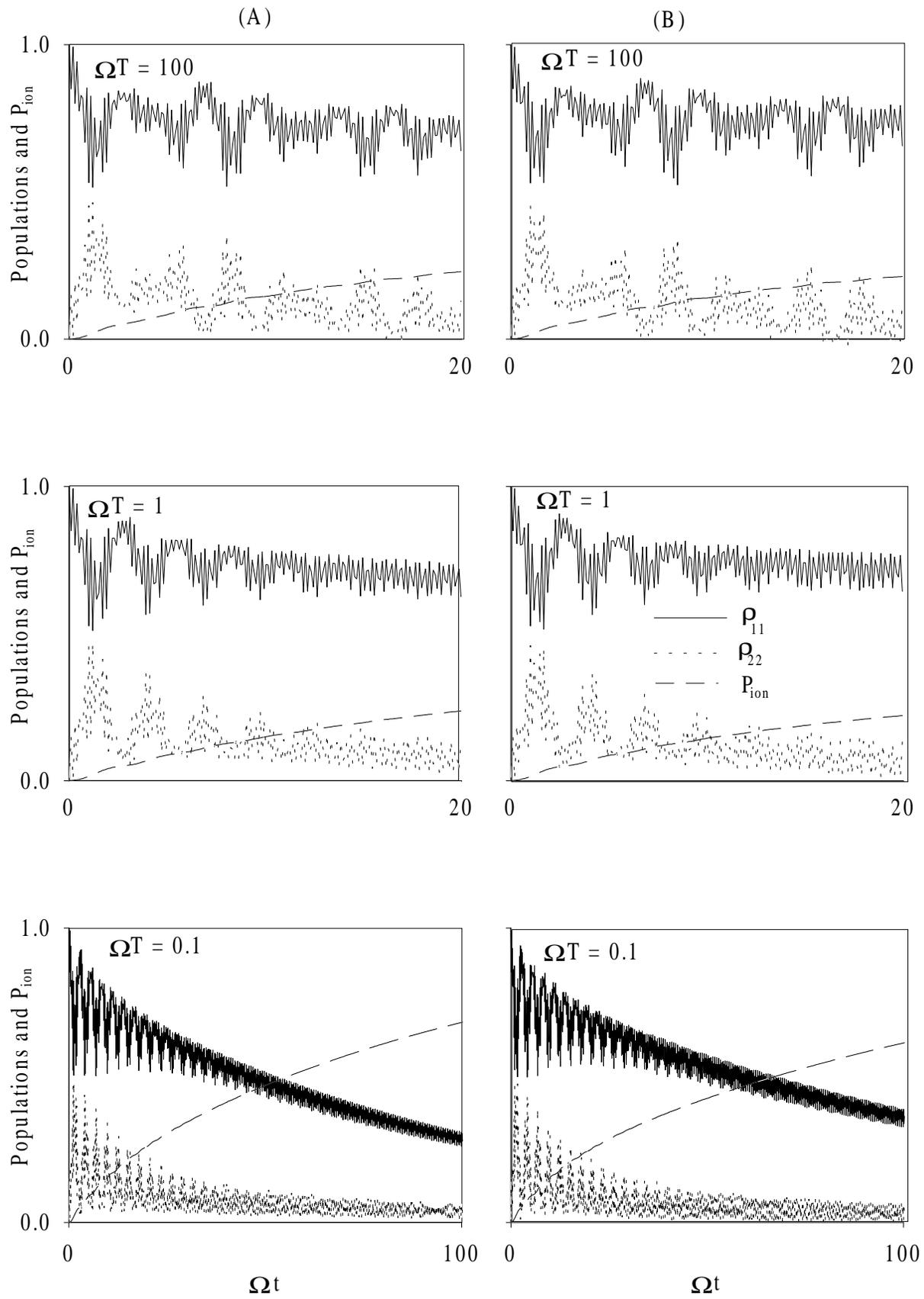

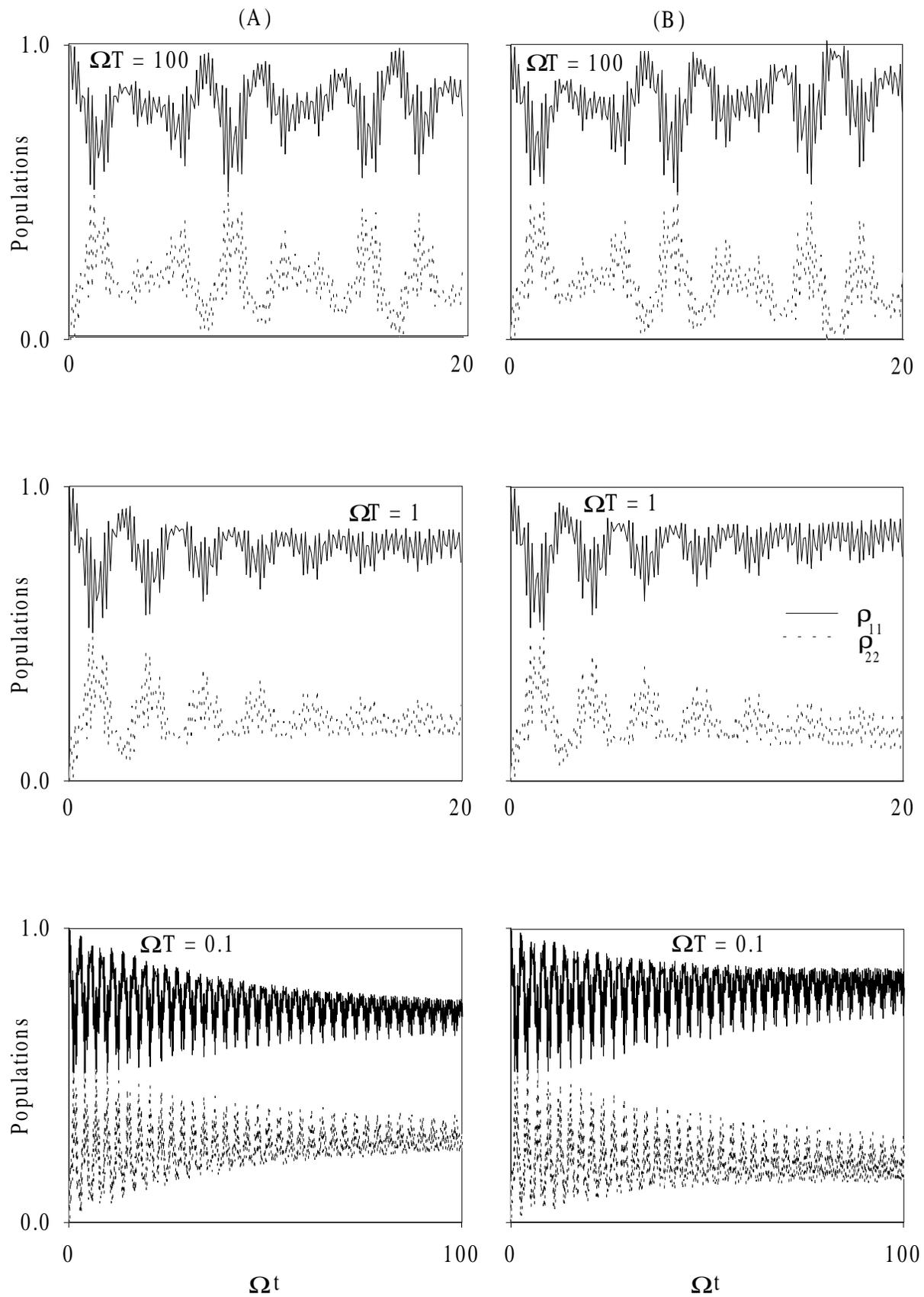

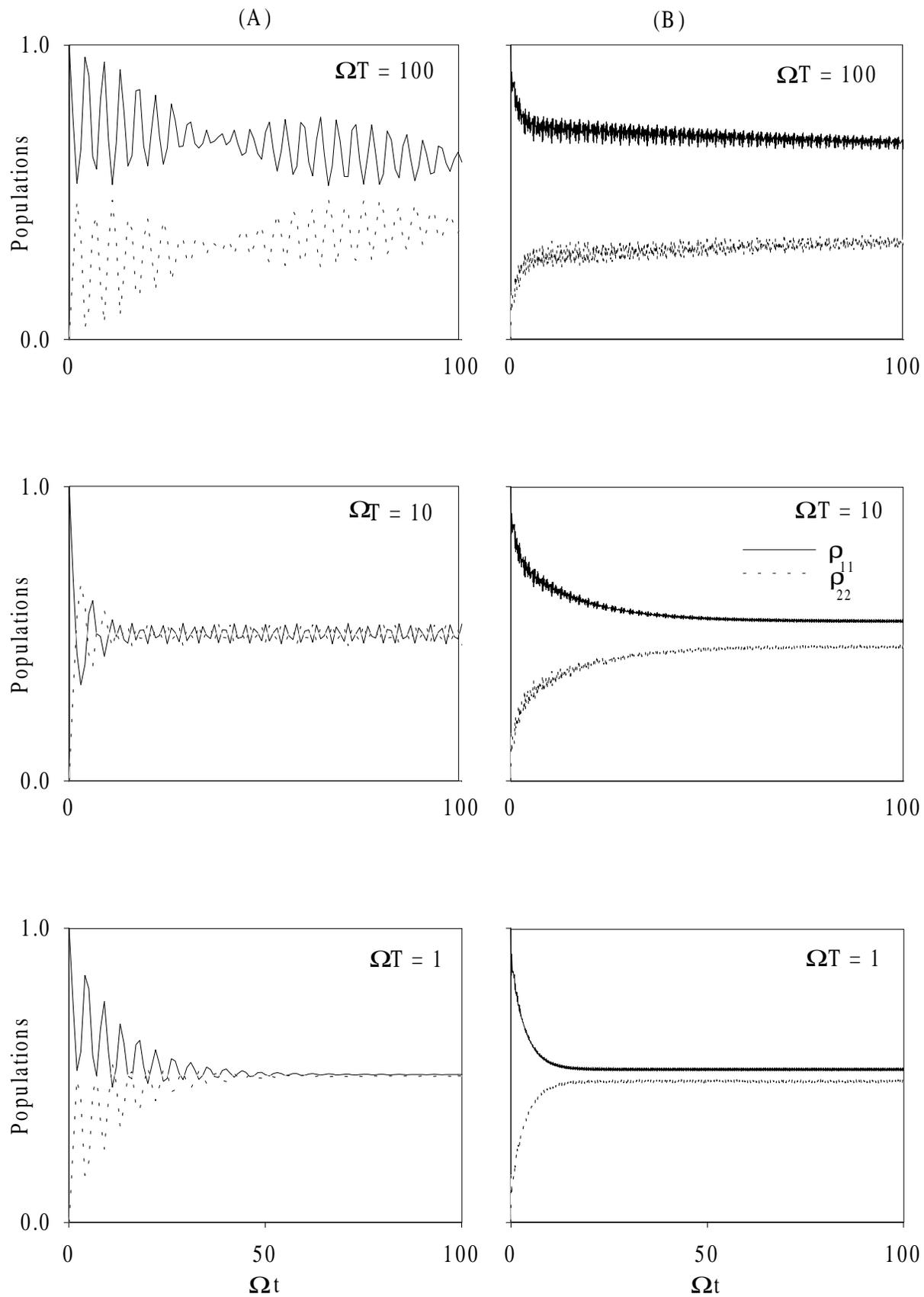

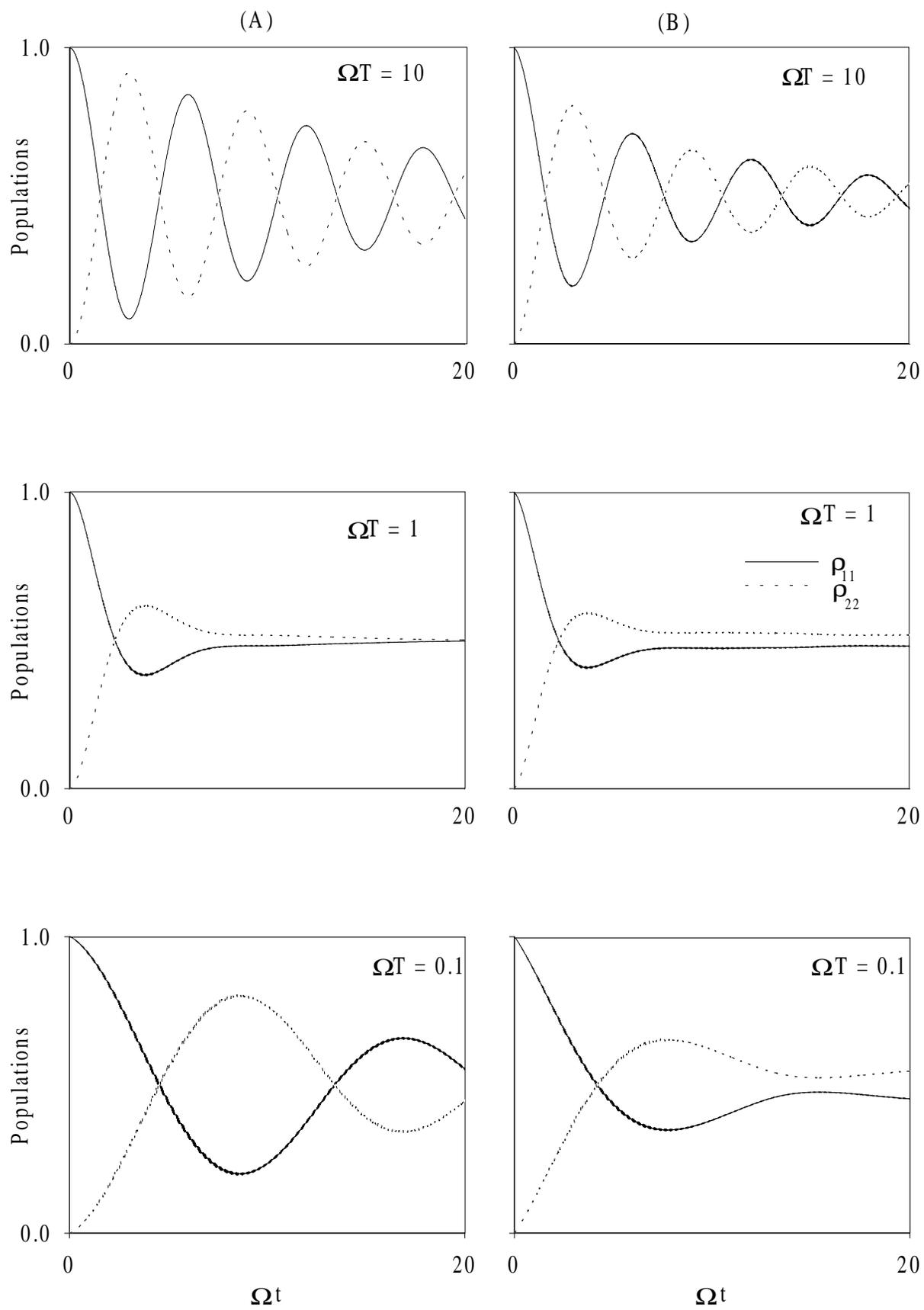

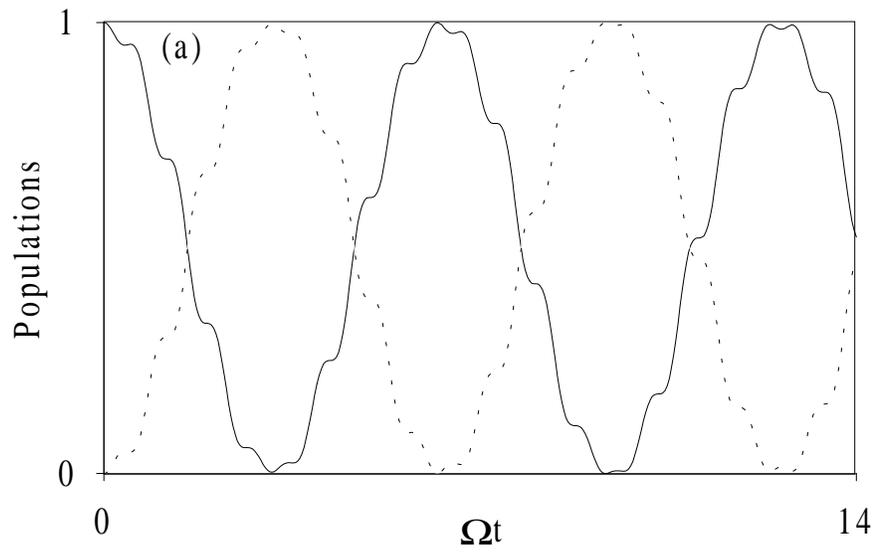

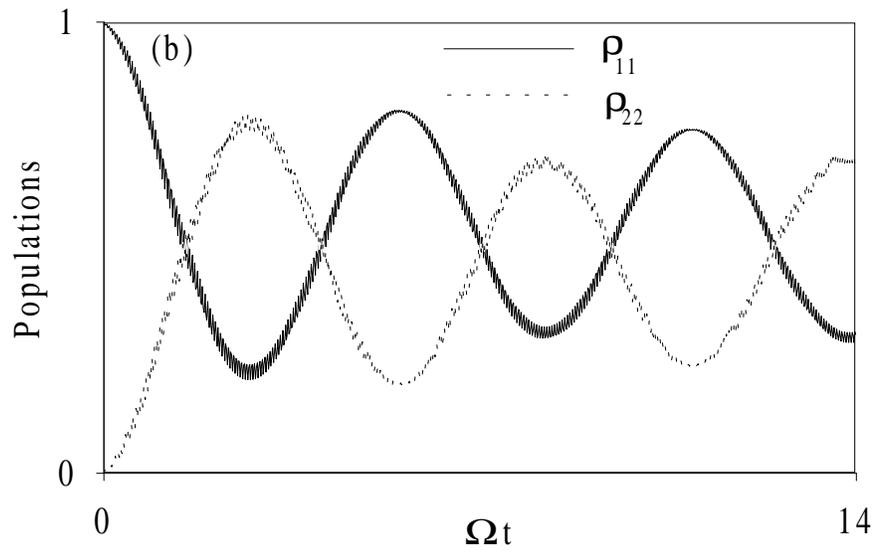

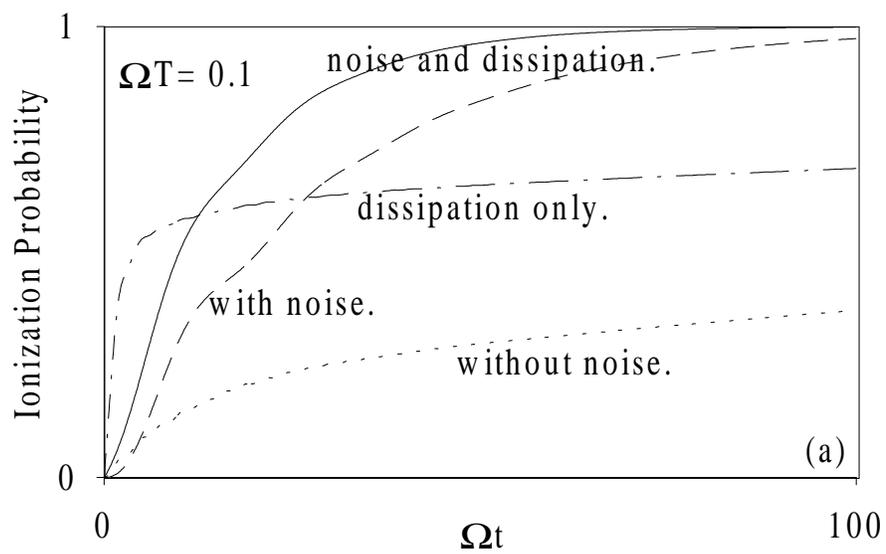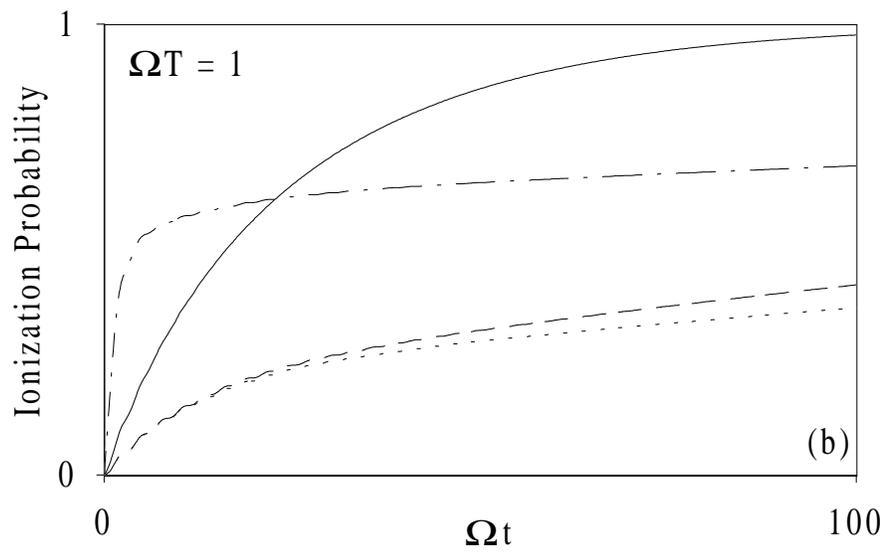